\documentclass[final,5p,times,twocolumn]{elsarticle}
\usepackage{amssymb}
\usepackage{lipsum}

\biboptions{numbers,sort&compress}
\setcitestyle{square,comma}
\usepackage[colorlinks=true,linkcolor=blue]{hyperref}%
\usepackage{physics}
\usepackage{graphicx}
\usepackage{lipsum}
\usepackage{mwe}
\usepackage{caption}
\usepackage{cancel}
\usepackage{todonotes}
\usepackage[normalem]{ulem}

\usepackage{dcolumn}
\usepackage{bm}
\usepackage{adjustbox}
\usepackage{rotating}
\usepackage{xcolor}
\usepackage{ulem}
\usepackage{braket}
\usepackage[dvipsnames]{xcolor}

\newcommand\MDsout{\bgroup\markoverwith{\textcolor{RubineRed}{\rule[0.5ex]{2pt}{0.7pt}}}\ULon}

\journal{Physics Letters B}

\begin{document}

\begin{frontmatter}

\title{Density-driven $\alpha$-particle clustering in neutron-rich oxygen isotopes}

\author[surrey]{M. Davies}
\ead{miriam.davies@surrey.ac.uk}
\author[surrey]{E. Y\"{u}ksel}
\ead{e.yuksel@surrey.ac.uk}
\author[CEA,Paris-Saclay]{J.-P. Ebran}
\author[IJClab,IUF]{E. Khan}

\affiliation[surrey]{organization={School of Mathematics and Physics, University of Surrey},
            city={Guildford},
            postcode={GU2 7XH},
            country={United Kingdom}}

\affiliation[CEA]{organization={CEA,DAM,DIF},
            city={F-91297 Arpajon},
            country={France}}

\affiliation[Paris-Saclay]{organization={Universit\'{e} Paris-Saclay, CEA, Laboratoire Mati\`{e}re en Conditions Extr\^{e}mes},
            city={91680, Bruy\`eres-le-Ch\^{a}tel},
            country={France}}

\affiliation[IJClab]{organization={IJCLab, Universit\'e Paris-Saclay, CNRS/IN2P3},
            city={91405 Orsay Cedex},
            country={France}}

\affiliation[IUF]{organization={Institut Universitaire de France (IUF)}}

\begin{abstract}
In this work, we study the effect of valence neutrons on
density-driven localisation in the neutron-rich oxygen isotopes. Calculations are performed within the
multidimensionally constrained relativistic Hartree--Bogoliubov model using the DD-ME2 covariant energy-density functional. By
constraining the mass rms radius to larger values, we study the structural evolution of these nuclei under progressive dilution, from spatially delocalised configurations to localised clustered states. The total energy, pairing correlations, deformation, and proton and neutron density distributions are analysed to determine how valence neutrons influence the emergence of $\alpha$-like clustering. For all three isotopes, we find a transition from a spatially delocalised nucleonic configuration to a tetrahedral $4\alpha$-like structure. The onset of this transition occurs at similar values of the radius-derived relative dilution parameter, $\chi_{\mathrm{PT}}=\rho_{\mathrm{PT}}/ \rho_{\mathrm{GS}}\simeq 0.28$--$0.31$, and shows only a weak dependence on neutron number. The excess neutrons modify the
clustered density by increasing the density in the central regions and between clusters, while leaving the underlying tetrahedral core structure largely intact. Finally, we also use a phenomenological four-centre H\"uckel-type mapping of nuclei, which provides a molecular-orbital-like interpretation of cluster structures. We find that the effective coupling between the cluster centers decreases by progressively expanding the nucleus. The centroid-referenced energy shift $\Delta E_{\mathrm{val}}$ also becomes less negative and provides smaller net molecular-orbital-like energy lowering relative to its centroid with increasing neutron number.

\end{abstract}


\end{frontmatter}

\section{\label{sect:intro}Introduction}

Clustering in nuclei -- where nucleons localise into spatially correlated substructures -- was initially predicted in the 1930s and is still one of the most interesting structural phenomena in nuclear many-body systems \cite{v1938neuere,wheeler1937mathematical}. These localised structures are usually associated with $\alpha$-like clusters and are particularly prominent in light, self-conjugate ($N=Z$) nuclei \cite{Clusters_in_nuclei,Ebran2012,PhysRevC.87.044307,Ebran2017, Freer_2007}. A well-known example of $\alpha$ clustering is the $0^+_2$ Hoyle state in $^{12}$C \cite{hoyle1954nuclear}, which facilitates the triple-$\alpha$ process and consequently, carbon synthesis in stars \cite{FREER20141, RevModPhys.89.011002}. The idea that $^{16}$O has a tetrahedral $4\alpha$ structure has also been proposed widely \cite{BRINK1970143, PhysRevLett.42.876, PhysRevC.25.1108}; the algebraic cluster model has provided evidence for tetrahedral $4\alpha$ symmetry in the ground-state band of $^{16}$O \cite{PhysRevLett.112.152501}, and the \textit{ab initio} nuclear lattice effective field theory calculations also find a predominantly tetrahedral arrangement of $\alpha$ clusters in the ground state of $^{16}$O \cite{PhysRevLett.112.102501}. Further to this, microscopic four-body orthogonality condition model (OCM) calculations relate the experimental $0^+_6$ state at $15.1$~MeV to a dilute and condensate-like $4\alpha$ configuration \cite{16O-hoyle-analogue}, and recently, four narrow resonances immediately above this state were observed in the $\alpha+{}^{12}$C Hoyle-state decay channel, providing additional evidence for a possible Hoyle-like $4\alpha$ structure above the $4\alpha$ separation threshold \cite{CHEN20231119}.

Although the most prominent examples of $\alpha$ clustering occur in light nuclei along the $N=Z$ line, it is known that cluster correlations can also persist in neutron-rich systems. In such nuclei, valence neutrons occupy molecular orbitals that extend over cluster cores and enhance the stability of the underlying cluster configurations \cite{VONOERTZEN1997, VONOERTZEN2003465, VONOERTZEN200643, Freer_2007}. For example, $^{8}$Be is unbound with respect to breakup into two $\alpha$ particles, but the addition of a neutron produces $^{9}$Be, which is bound with respect to the $2\alpha+n$ threshold. Here, the valence neutron provides covalent-like binding between the two $\alpha$ clusters \cite{VONOERTZEN1996}. Similarly, in $^{10}$Be, two valence neutrons result in a rich spectrum of $2\alpha+2n$ molecular configurations \cite{VONOERTZEN1996}. Cluster structures have also been predicted in the neutron-rich oxygen isotopes $^{18}$O and $^{20}$O \cite{10.1143/PTP.119.403}.

Experimental studies have provided some evidence for clusters and molecular structures in neutron-rich nuclei. Breakup studies in neutron-rich beryllium and carbon isotopes have identified resonant states via cluster-decay channels \cite{PhysRevC.93.024611}. Resonant-scattering experiments involving $^{13}$B and $^{14}$C have also revealed further structures in cluster channels, whereas their interpretation as evidence for exotic clustering remains unresolved \cite{Di_Pietro_2018}. Recent inelastic excitation and cluster-decay experiments on $^{16}$C identified a sequence of resonances which is consistent with a positive-parity linear-chain molecular band \cite{PhysRevLett.124.192501}, which also supports theoretical predictions given in Refs. \cite{PhysRevC.90.064319, PhysRevC.97.054315, PhysRevC.64.014301}. On the other hand, breakup measurements of $^{16}$C did not reveal a well-developed cluster structure in its ground state and may indicate the pronounced state dependence of clustering phenomena in neutron-rich systems \cite{PhysRevC.70.064607}.  Experiments also support $\alpha$-cluster and molecular configurations in excited $^{18}$O~\cite{PhysRevC.90.024327,PhysRevC.99.064315}. Oxygen isotopes thus offer a natural setting in which to study how neutron excess affects density-driven clustering at fixed $Z=8$.

Self-consistent calculations have also shown that $^{16}$O evolves from a more spatially delocalised nucleonic configuration to an $\alpha$ clustered structure as the density is reduced under quadrupole-suppressed radial expansion \cite{GIROD-clustering, PRC-16O_QPT,DAVIES-16O-PT-FT}. In infinite nuclear matter, the conventional Mott effect refers to the dissolution of bound clusters with increasing density as a result of in-medium effects, in particular Pauli blocking \cite{BEYER2000247, PhysRevC.81.015803, Ropke2015}. The finite nucleus phenomenon considered here is complementary; decreasing the average density under constrained expansion drives a structural reorganisation in nuclei towards localised $\alpha$-like configurations. Previously, this density-driven reorganisation has been interpreted as a Mott-like transition occurring near a critical sub-saturation density \cite{GIROD-clustering, PRC-16O_QPT}. Although clustering and molecular-orbital structures in neutron-rich light nuclei have been extensively studied over the years, the density-driven emergence of clustering in dilute neutron-rich finite nuclei remains poorly understood. In particular, it remains unclear how excess neutrons, through their single-particle occupations, pairing correlations, and molecular-orbital bonding, influence the Mott-like transition toward localisation in radially expanded neutron-rich oxygen isotopes.

In this work, we study localisation and clustering in neutron-rich oxygen isotopes $^{18,20,24}$O, using $^{16}$O as a reference system, within the multidimensionally constrained relativistic Hartree--Bogoliubov (MDC-RHB) framework and using the DD-ME2 energy-density functional. By following the evolution of the total energy, pairing correlations, deformation, and proton and neutron densities, we study how excess neutrons affect the formation and stability of dilute $4\alpha$-like configurations under constrained isotropic expansion. Lastly, to assess the role of the valence neutrons in the clustered regime, we also employ a four-centre H\"uckel-type mapping of the relevant RHB single-particle spectrum.

\section{Theoretical formalism}\label{sect:theory}
Relativistic Hartree-Bogoliubov (RHB) theory combines self-consistent mean-field dynamics with pairing correlations and provides a microscopic framework for calculating nuclear ground-state properties. Over the years, it has been successfully applied to calculate the ground-state properties of nuclei (see Refs.~\cite{ring1980nuclear,VRETENAR2005101,dirhb, DAVIES-16O-PT-FT} for further details). In this study, we use the multidimensionally constrained relativistic Hartree-Bogoliubov (MDC-RHB) framework with the density-dependent meson-exchange functional DD-ME2 \cite{PhysRevC.71.024312} and the finite-range separable pairing interaction \cite{TIAN200944,dirhb}.  We performed our calculations by imposing constraints on mass multipole moments and the mass rms radius, $R_{\mathrm{rms}}$, the latter of which acts as a collective radial coordinate that controls the overall spatial extension of the nucleonic density and the degree of dilution. In the quadratic constraint method, we minimise the RHB energy functional with an extra quadratic constraint term

\begin{equation}
E'_{\rm RHB}
=
E_{\rm RHB}
+
\sum_i C_i
\left(
\langle \hat{Q}_i \rangle
-
q_i
\right)^2 ,
\label{eqn:quadratic_constraint}
\end{equation}
where $E_{\rm RHB}$ denotes the total RHB energy, $\hat{Q}_i$ is a collective operator, $q_i$ is its prescribed expectation value, and $C_i$ is the corresponding stiffness constant that controls the strength of the constraint during iterative minimisation. To improve numerical stability and ensure convergence towards prescribed collective coordinates, we implement the constraints using the augmented Lagrangian method. \cite{Staszczak2010,dirhb}. The collective operators for the mass multipole constraints are defined as

\begin{equation}
\hat{Q}_{\lambda\mu}
=
\sum_{a=1}^{A}
r_a^\lambda
Y_{\lambda\mu}(\Omega_a),
\label{eqn:multipole_operator}
\end{equation}
where $r_a$ and $\Omega_a$ denote the radial and angular coordinates of nucleon $a$, respectively. For reflection-asymmetric configurations, the centre of mass is fixed at the origin by imposing the additional constraint $\langle \hat{Q}_{10} \rangle = 0$. This removes spurious dipole contributions associated with translations of the intrinsic density. The dimensionless mass deformation parameters are defined as

\begin{equation}
\beta_{\lambda \mu}
=
\frac{4\pi}{3 A R_{0}^{\lambda}}
\langle \hat{Q}_{\lambda \mu} \rangle ,
\label{eqn:beta_lm}
\end{equation}
where $R_{0}=r_{0}A^{1/3}$ with $r_{0}=1.2$ fm. Analogous neutron and proton deformation parameters are obtained by replacing $A$ with $N$ or $Z$, respectively, and by using the corresponding neutron or proton multipole moments. Since the dilute configurations considered here are spatially extended, the harmonic-oscillator (HO) basis size and oscillator frequency, $\hbar\omega$, must be chosen carefully to minimise basis-truncation artefacts. Previous studies of dilute $^{16}$O have conducted similar basis optimisations \cite{PRC-16O_QPT, DAVIES-16O-PT-FT}. In this study, we find that using $N_{\mathrm{shell}}=12$ HO shells and $\hbar\omega=13$ MeV provides stable and well-converged solutions with properly minimised energies over the range of radial constraints considered in this study. The optimal basis parameters may depend on both the mass number and the spatial extension of the isotope; therefore, we also performed additional convergence checks to ensure that these choices remain reliable for the neutron-rich oxygen isotopes studied here.

\section{Results and discussion}\label{sect:results}

\begin{figure*}[htb!]
\centering
\begin{minipage}{0.45\linewidth}
    \centering
    \includegraphics[width=\linewidth]{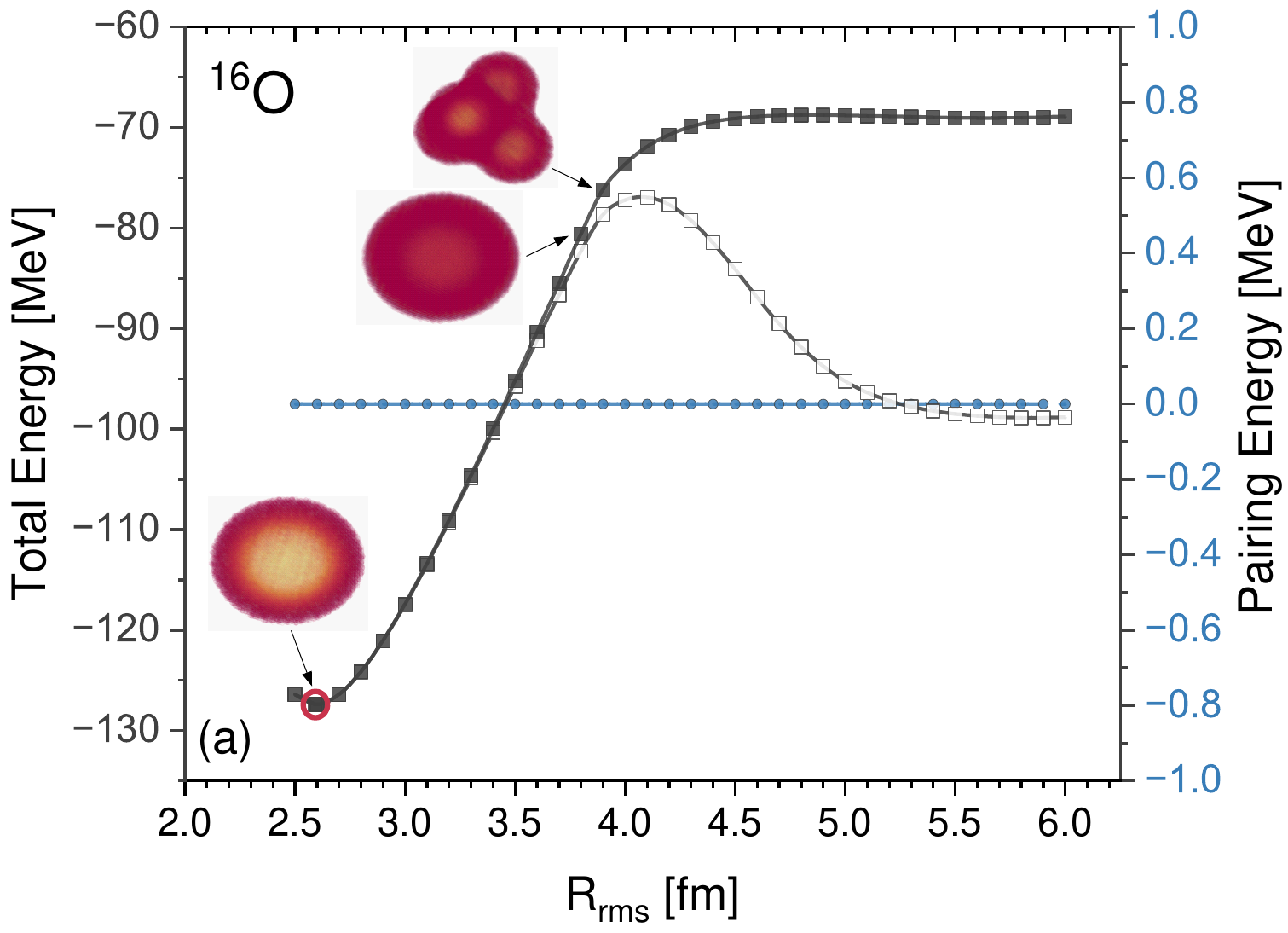}
\end{minipage}\hspace{0.01\linewidth}
\begin{minipage}{0.45\linewidth}
    \centering
    \includegraphics[width=\linewidth]{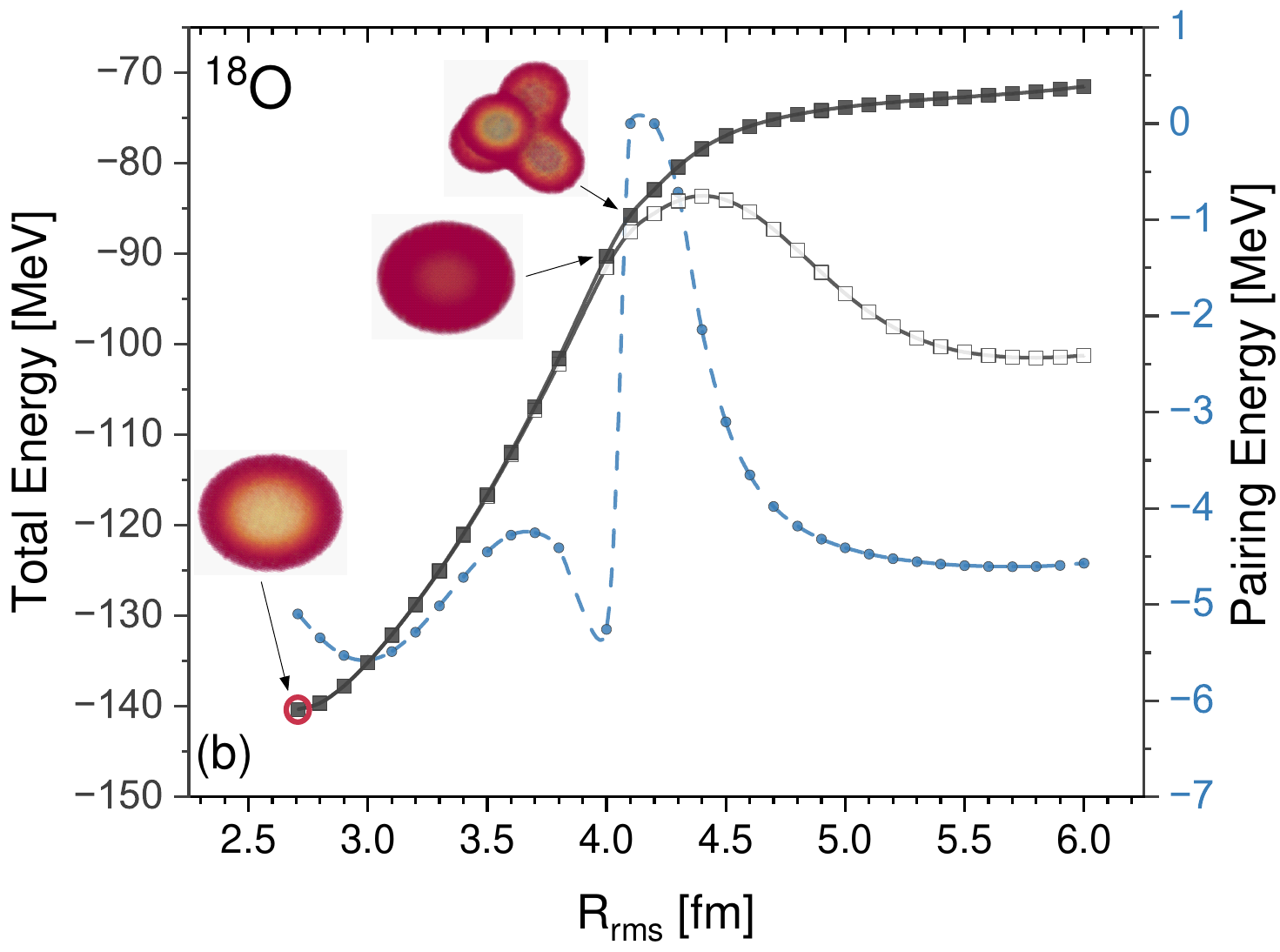}
\end{minipage}
\vspace{0.05cm}
\begin{minipage}{0.45\linewidth}
    \centering
    \includegraphics[width=\linewidth]{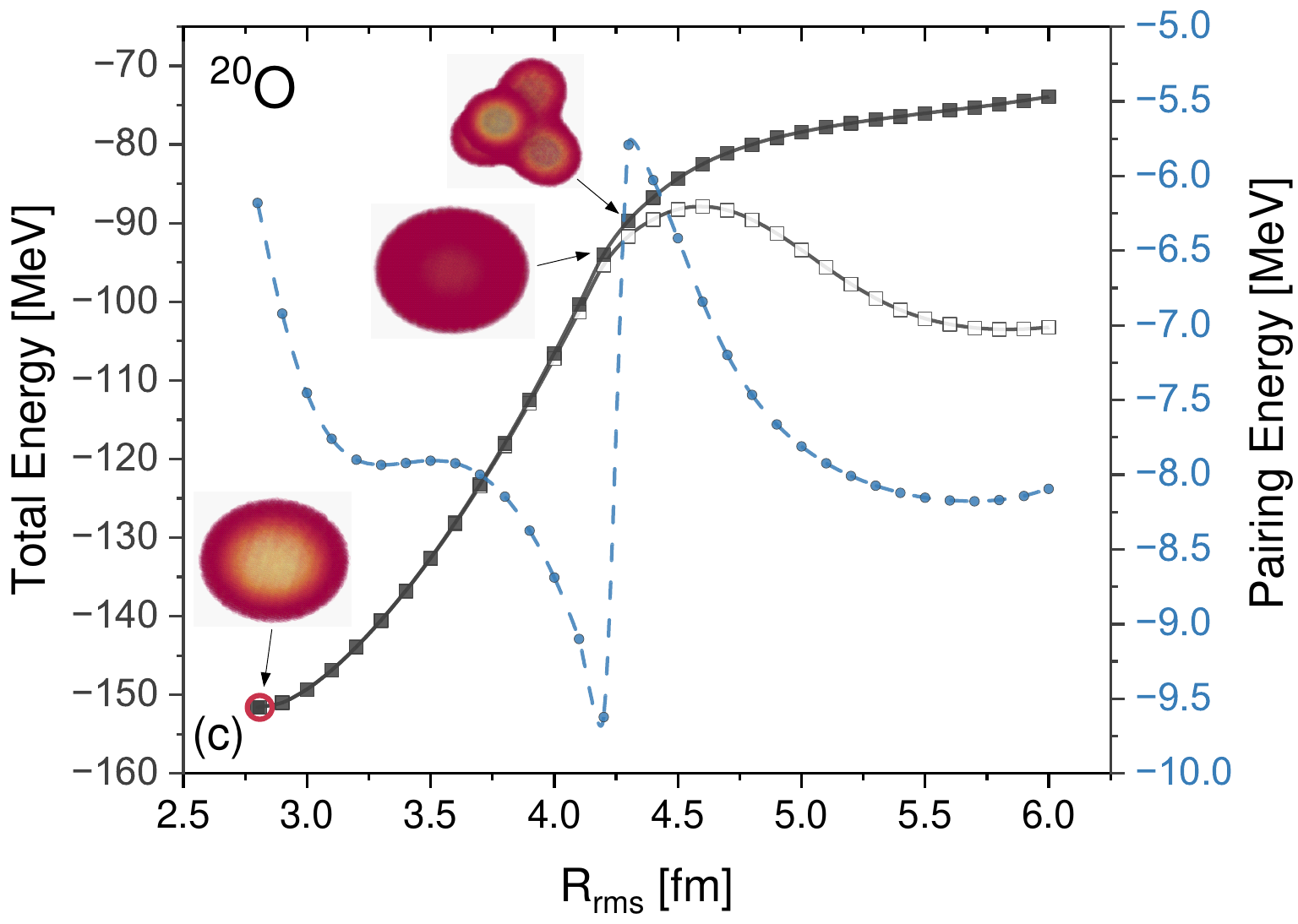}
\end{minipage}\hspace{0.01\linewidth}
\begin{minipage}{0.45\linewidth}
    \centering
    \includegraphics[width=\linewidth]{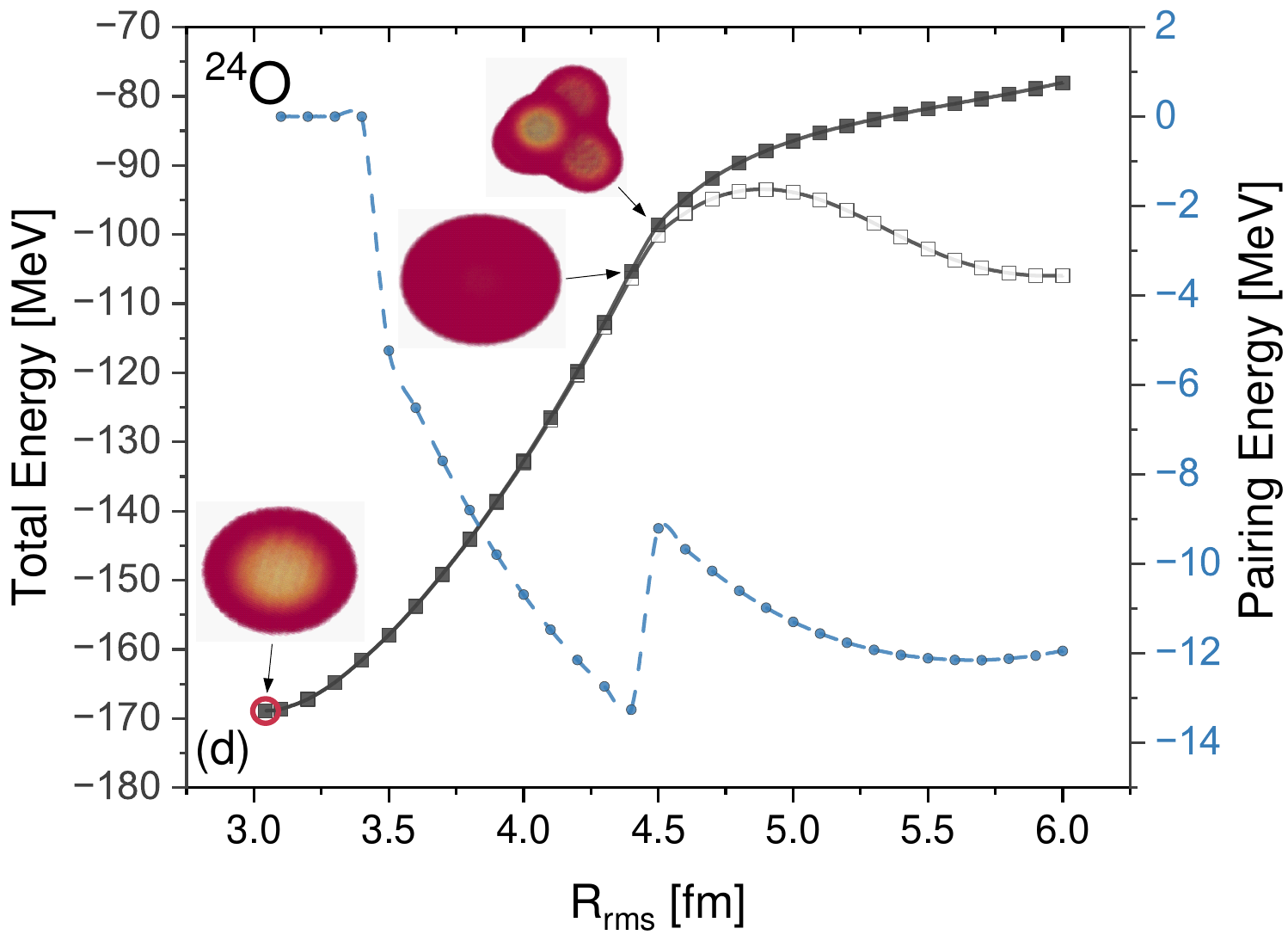}
\end{minipage}
\caption{
Total energy and pairing energy as functions of the constrained
$R_{\mathrm{rms}}$ for 
(a) $^{16}$O, (b) $^{18}$O, (c) $^{20}$O, and (d) $^{24}$O. 
The total energy is shown by the solid grey squares and plotted on the left-hand axis, while the pairing energy is shown by blue circles and plotted on the right-hand axis. The ZPE corrected total energy is given on the left-hand axis as open grey squares. For each isotope, the ground-state configuration is highlighted by a red circle. Three-dimensional density distributions, shown with an isosurface value of $\rho_{\mathrm{min}}=0.005$ fm$^{-3}$, are displayed at selected radii corresponding to the ground state, just before the transition, and just after the transition, in order to illustrate the associated structural change.
}
\label{fig:Etot-Pairing-O16-O18-O20-O24}
\end{figure*}

The localisation and formation of $\alpha$ cluster structures in dilute $^{16}$O have previously been studied by progressively expanding the nucleus, i.e., by constraining its mass rms radius to increasingly large values. Under progressive dilution, $^{16}$O has been shown to undergo a Mott-like transition from a relatively homogeneous, delocalised configuration to a clustered, localised structure \cite{PRC-16O_QPT,GIROD-clustering,DAVIES-16O-PT-FT}. Related radius-constrained HFB calculations using the Gogny interaction predict spontaneous $\alpha$ clustering in several self-conjugate nuclei at average densities of approximately one third of their equilibrium densities \cite{GIROD-clustering}. Studies in symmetric nuclear matter also predict cluster formation and, at sufficiently low temperatures, $\alpha$-like four-particle condensation below a critical subsaturation density, $\rho_{\mathrm{c}}<\rho_{\mathrm{sat}}$, where $\rho_{\mathrm{sat}}\approx0.16$~fm$^{-3}$ \cite{ROPKE1982536,PhysRevLett.80.3177}. In this study, we examine neutron-rich oxygen isotopes to understand how valence neutrons affect density-driven localisation and whether they help or prevent cluster formation.

In Figure~\ref{fig:Etot-Pairing-O16-O18-O20-O24}, we show the total energy, displayed on the left-hand axis, and the pairing energy, displayed on the right-hand axis, as a function of the constrained $R_{\mathrm{rms}}$ for $^{16}$O, $^{18}$O, $^{20}$O, and $^{24}$O in panels (a)--(d), respectively. To expand nuclei, the nuclear $R_{\mathrm{rms}}$ is gradually increased in the calculations, and global quadrupole deformation is kept suppressed by maintaining $\langle Q_{2\mu}\rangle=0$ ($\beta_{2\mu}=0$) during the entire process. This isolates the radius-driven structural evolution and allows an isotropic nuclear inflation. We also note that the ground states of the oxygen isotopes considered have zero quadrupole deformation. All other symmetry-allowed mass multipole moments, i.e., the axial octupole deformation $\beta_{30}$, the non-axial octupole deformation $\beta_{32}$, and the axial hexadecapole deformation $\beta_{40}$ are left free and unconstrained. This allows the nuclei to develop reflection-asymmetric, non-axial, or higher-order deformations if they are energetically favourable. Here, we should mention that the non-axial octupole component $\beta_{32}$ is particularly important as it provides the leading degree of freedom associated with tetrahedral cluster configurations. Pairing correlations are included self-consistently and also monitored along the radial constraint as a diagnostic of quasiparticle rearrangement. No pairing correlations are obtained in $^{16}$O over the range of radial constraints considered. At the ground state, pairing also vanishes in the closed-neutron-shell nucleus $^{24}$O, whereas we obtain neutron pairing in open-shell $^{18}$O and $^{20}$O nuclei. During expansion, neutron pairing develops in $^{24}$O as the single-particle spectrum rearranges.

The behaviour of $^{16}$O is considered first as a reference case for the density-induced transition in an $N=Z$ oxygen isotope. Figure~\ref{fig:Etot-Pairing-O16-O18-O20-O24}(a) shows the total-energy curve, which is consistent with previous studies of dilute $^{16}$O \cite{PRC-16O_QPT,GIROD-clustering,DAVIES-16O-PT-FT}. Open grey squares show the energies obtained after applying a phenomenological correction for the spurious centre-of-mass zero-point energy of the spatially localised $\alpha$-like fragments \cite{DAVIES-16O-PT-FT,PRC-16O_QPT}. Details of this correction are given in the Supplemental Material. As the constrained $R_{\mathrm{rms}}$ increases, the total energy of $^{16}$O increases rapidly, namely, the nucleus becomes less bound, before approaching a plateau at $R_{\mathrm{rms}}\gtrsim4.4$~fm and changes only weakly under further expansion. This stabilisation is accompanied by a sudden change to a non-zero $\beta_{32}$ and a shift toward a tetrahedral structure. Rather than continuing to dilute smoothly, the density localises into four spatially separated regions with local densities close to saturation density, corresponding to the formation of a $4\alpha$-like structure. With further expansion, the four $\alpha$-like substructures become spatially more separated.

\begin{figure}
	\centering 
	\includegraphics[width=0.9\linewidth]{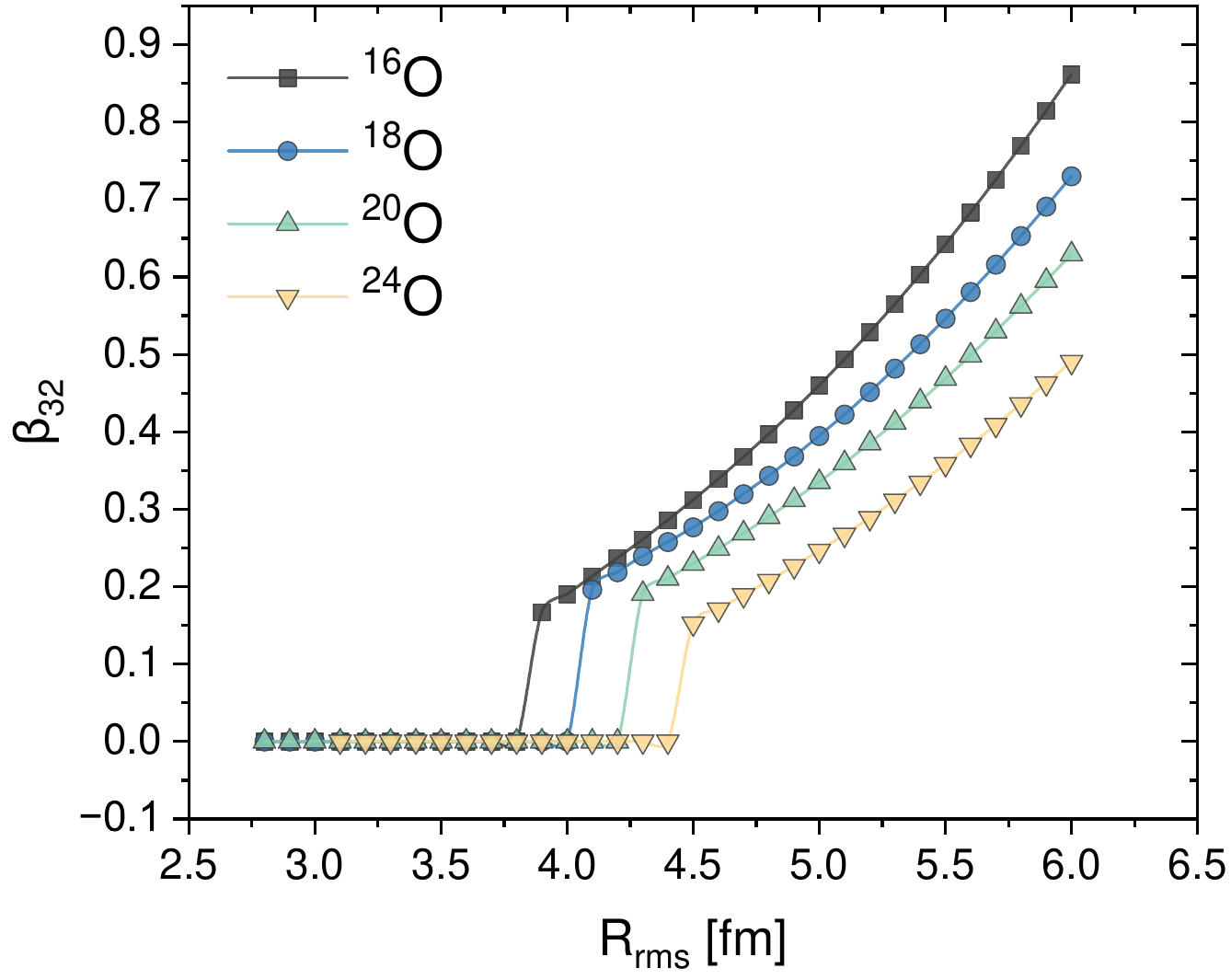}	
	\caption{The evolution of the non-axial octupole deformation, $\beta_{32}$, in $^{16}$O, $^{18}$O, $^{20}$O, and $^{24}$O, as a function of increasing radial constraint.} 
	\label{fig:b32-O16-O18-O20-O24}%
\end{figure}

We next study the neutron-rich isotopes $^{18}$O, $^{20}$O, and $^{24}$O. As shown in Figs.~\ref{fig:Etot-Pairing-O16-O18-O20-O24}(b)--(d), these nuclei show qualitatively similar total energy evolution under radial expansion, including the development of tetrahedral localisation. The total energy increases rapidly with increasing mass $R_{\mathrm{rms}}$, then varies more gradually after the onset of tetrahedral localisation. In contrast, the behaviour of pairing energy varies depending on the specific nucleus under consideration. For $^{16}$O, we do not see any pairing correlations across the range of radial constraints studied. For neutron-rich oxygen isotopes, only neutron pairing plays a role, while proton pairing remains zero during the expansion. This happens because the proton shell is closed at $Z=8$, and the extra neutrons in $^{18}$O, $^{20}$O, and $^{24}$O fill states above the $N=8$ shell closure. Furthermore, the pairing energy changes non-monotonically in the neutron-rich isotopes as the system expands.  In each case, there is a sudden change in the pairing energy at an isotope-dependent radius, which marks a transition region where the valence neutron single-particle spectrum and occupations rearrange rapidly. This region also coincides with the change in the slope of the total energy curve and with the development of a non-zero $\beta_{32}$ deformation, as shown in Fig.~\ref{fig:b32-O16-O18-O20-O24}. To assess the functional dependence of our results, we repeated the $^{20}$O calculation using DD-PC1 functional \cite{PhysRevC.78.034318}. Both DD-ME2 and DD-PC1 predict the same qualitative transition to a tetrahedral $4\alpha$-like configuration.

Beyond the transition region, and at larger constrained $R_{\mathrm{rms}}$, the pairing energy changes more gradually and varies only weakly. This shows that the quasiparticle structure becomes less sensitive to further radial expansion after the onset of tetrahedral localisation. In this situation, further expansion mostly increases the distance between $\alpha$-like substructures, while the pairing correlations of valence neutrons change more gradually. We interpret the sharp changes in pairing energy as signatures of quasiparticle rearrangement linked to the onset of tetrahedral deformation, not as evidence of clustering. Therefore, we also examine density distributions at selected points along the constrained energy curves. Three-dimensional density distributions with an isosurface value of $\rho_{\mathrm{min}}=0.005$~fm$^{-3}$ are also shown in Fig.~\ref{fig:Etot-Pairing-O16-O18-O20-O24} for the ground-state configurations and for configurations just before and after the transition. These densities clearly show the emergence of localisation and $\alpha$-like cluster structures as the systems are expanded.

Figure~\ref{fig:b32-O16-O18-O20-O24} shows the evolution of $\beta_{32}$ during radial expansion. For every isotope, $\beta_{32}$ stays at zero from the ground state up to moderately expanded configurations. It then changes suddenly once it reaches a certain critical $R_{\mathrm{rms}}$. This sudden onset provides a clear indicator of the transition region associated with tetrahedral localisation. The figure shows that onset of non-zero $\beta_{32}$ is shifted to larger $R_{\mathrm{rms}}$ as neutrons are added, and this shows that excess neutrons delay, but do not suppress, the development of tetrahedral deformation.

Lastly, we also introduce the relative dilution parameter $\chi_{\mathrm{PT}}$ to compare localisation points in isotopes that have different ground-state radii. By assuming an approximately isotropic radial expansion, this quantity is defined as
\begin{equation}
    \chi_{\mathrm{PT}}
    =
    \frac{\rho_{\mathrm{PT}}}
    {\rho_{\mathrm{GS}}}
    \simeq
    \left(
    \frac{R_{\mathrm{GS}}}{R_{\mathrm{PT}}}
    \right)^3 ,
    \label{eqn:norm_PT_density}
\end{equation}
where $\rho_{\mathrm{GS}}$ is the ground-state global reference density, $R_{\mathrm{GS}}$ is the ground-state mass rms radius, and $R_{\mathrm{PT}}$ is the mass rms radius at the transition point. The corresponding values are given in Table~\ref{tab:norm_PT_density}. Our results show that even though the transition radius increases as the neutron number increases, $\chi_{\mathrm{PT}}$ remains within a narrow range of 0.278 to 0.309 across the isotopic chain. This suggests that tetrahedral localisation mainly depends on how diluted the expanded system is, not just on the mass number. Further validation of the density diagnostics is provided in the Supplementary Material.

\begin{table}[tb]
\centering 
\caption{\label{tab:norm_PT_density}
The relative dilution parameter, $\chi_{\mathrm{PT}}$, for the oxygen isotopes considered.}
\begin{tabular}{ccccc} 
 \hline
 & Isotope & $N$ & $\chi_{\mathrm{PT}}$ & \\ 
 \hline
 & $^{16}$O & 8  & 0.293 & \\
 & $^{18}$O & 10 & 0.288 & \\
 & $^{20}$O & 12 & 0.278 & \\
 & $^{24}$O & 16 & 0.309 & \\
 \hline
\end{tabular}
\end{table}

\begin{figure}[t!]
\centering
\includegraphics[width=\linewidth]{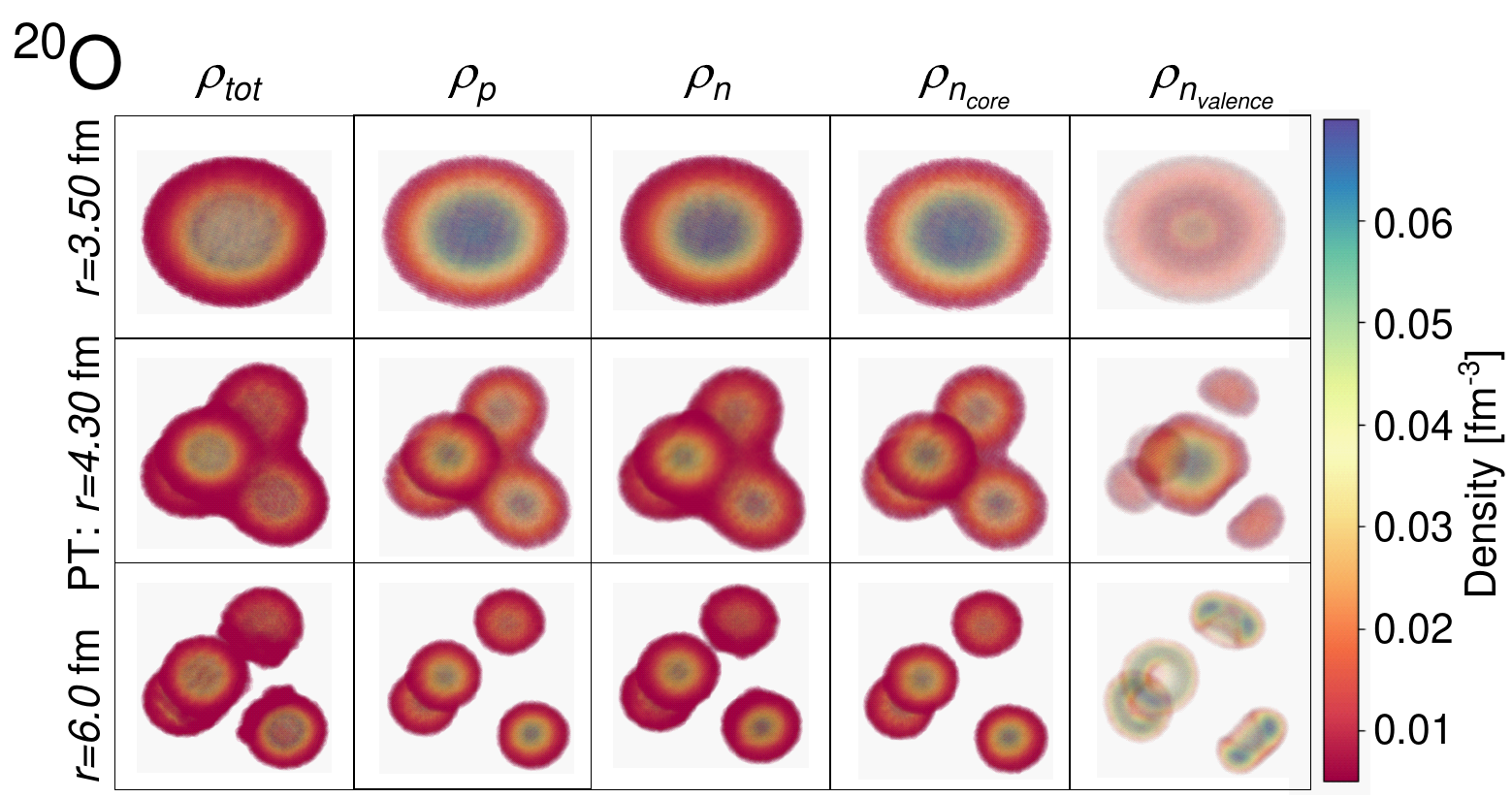}
\caption{
Three-dimensional constituent nucleon densities for $^{20}$O. The rows
correspond to three selected constrained rms radii: a compact configuration
at $R_{\mathrm{rms}}=3.50$~fm, the transition radius
$R_{\mathrm{PT}}=4.30$~fm, and an expanded configuration at
$R_{\mathrm{rms}}=6.0$~fm. The columns show the total density,
$\rho_{\mathrm{tot}}$, proton density, $\rho_{\mathrm{p}}$, neutron
density, $\rho_{\mathrm{n}}$, core-neutron density,
$\rho_{\mathrm{n,core}}$, and valence neutron density,
$\rho_{\mathrm{n,val}}$.
}
\label{fig:oxygen-density-grids}
\end{figure}

Figure~\ref{fig:oxygen-density-grids} displays the three-dimensional density distributions of $^{20}$O at three selected values of the mass rms radius: a relatively compact and delocalised configuration at $R_{\mathrm{rms}}=3.50$~fm, the transition configuration at $R_{\mathrm{PT}}=4.30$~fm, and an expanded and clustered configuration at $R_{\mathrm{rms}}=6.0$~fm. Also, the columns show the total density $\rho_{\mathrm{tot}}$, proton density $\rho_{\mathrm{p}}$, neutron density $\rho_{\mathrm{n}}$, core-neutron density $\rho_{\mathrm{n,core}}$, and valence neutron density $\rho_{\mathrm{n,val}}$, respectively. To distinguish the core and valence contributions, we decompose the neutron density into the partial densities of the selected RHB canonical orbitals. The core-neutron density is obtained from the lowest neutron orbitals associated with the $N=8$ neutron component of the $N=Z=8$, $^{16}$O-like core. The remaining neutron orbitals above the $N=8$ core define the valence neutron contribution, and we identify these orbitals by analysing the neutron canonical single-particle spectrum.

At $R_{\mathrm{rms}}=3.50$~fm, all the plots show that densities are smooth and spatially delocalised with no distinct cluster substructures. The valence neutron density is more diffuse and spatially extended than the core density, as expected. At the transition radius, the proton and core-neutron densities develop corresponding spatial maxima that form a tetrahedral $4\alpha$-like configuration. The similarity of these distributions shows that the strongest localisation occurs within the $N=Z=8$, $^{16}$O-like core. In comparison, it is observed that the valence neutron density remains comparatively diffuse and retains substantial density in the central region and between clusters, hence smooth the total neutron and total density distributions. With further expansion to $R_{\mathrm{rms}}=6.0$~fm, the $\alpha$-like substructures become more clearly separated. The valence neutron density remains more diffuse than the core density and extends around and between the individual clusters. Thus, we conclude that the valence neutrons modify and smooth the overall density distribution while preserving the the underlying tetrahedral $4\alpha$-like structure of the core.

\begin{figure}[htbp!]
\centering
\includegraphics[width=\linewidth]{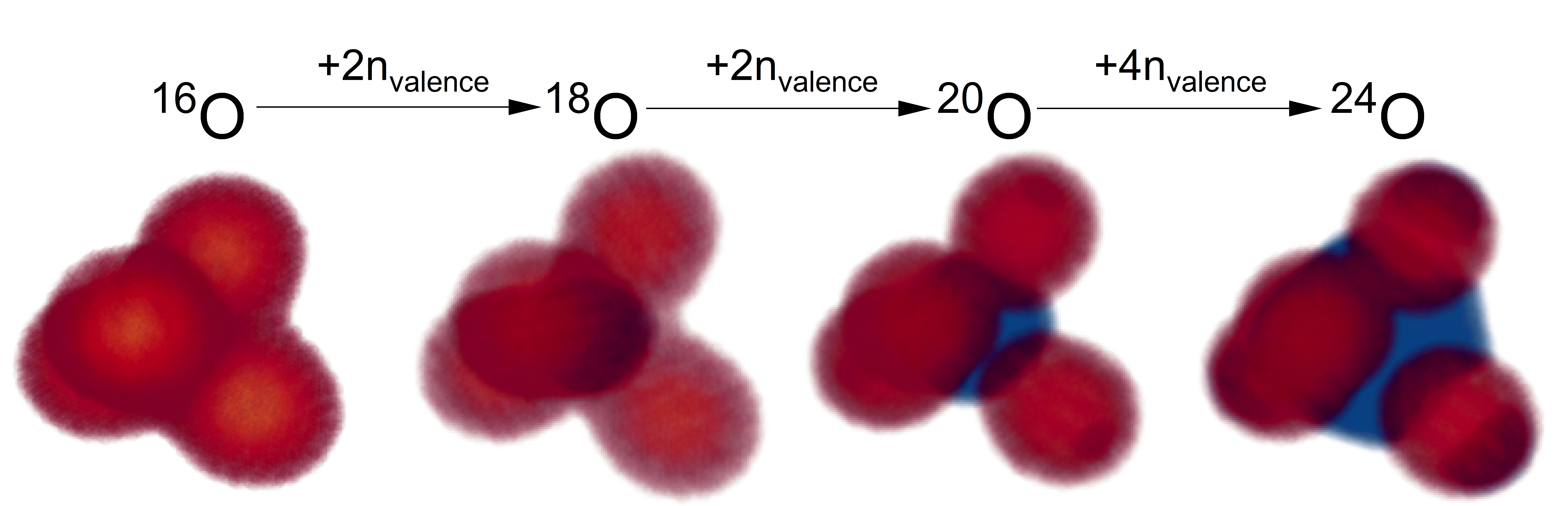}
\caption{Superpositions of the three-dimensional densities for the core neutrons (red) and valence neutrons (blue) for $^{16}$O, $^{18}$O, $^{20}$O, and $^{24}$O.}
\label{fig:superposition-O16-O18-O20-O24}
\end{figure}

Figure~\ref{fig:superposition-O16-O18-O20-O24} also illustrates how the tetrahedral $4\alpha$ structure of $^{16}$O changes as more valence neutrons are added. The red distribution represents the underlying core neutron density, and the blue distribution shows the valence neutron density. In $^{18}$O, the valence neutron density is not visible since it is relatively weak and largely overlaps with the core density at the chosen isosurface value. As the neutron number increases further, the valence neutron density becomes more pronounced in the central region and between the clusters, and extends across the clustered configuration. This effect is most pronounced in $^{24}$O, which has eight valence neutrons. Our results show that adding more neutrons does not destroy the $4\alpha$-like core. Instead, they increase the density between clusters and provide additional spatial connectivity between $\alpha$-like substructures. This is also in agreement with molecular-orbital descriptions in which valence neutrons are delocalised over a clustered core \cite{VONOERTZEN2003465,VONOERTZEN2004,PhysRevC.90.054329}.

Finally, to interpret the evolution of the valence neutron single-particle spectrum under constrained radial expansion and in terms of molecular-orbital-like states, we employ a phenomenological four-centre H\"uckel-type mapping \cite{Huckel1931,Huckel1932,VONOERTZEN200643,Brooks_2025}. This is inspired by molecular-orbital models in which valence nucleons occupy states extending over several cluster centres \cite{VONOERTZEN1997,VONOERTZEN2003465,VONOERTZEN200643,PhysRevC.64.014301}. We map a selected set of self-consistent RHB canonical states onto the one-plus-three-level structure of an ideal four-site tetrahedral model. This model contains one fully symmetric bonding-like state and three orthogonal upper partner states and provides a simple phenomenological description of the effective spectral splitting and occupation pattern of the selected valence neutron multiplet. It is important to note that we are not aiming to provide a microscopic derivation of localised cluster-centred orbitals or of molecular binding. In this study, the four sites represent the four $\alpha$-like density substructures in a phenomenological way. In the ideal tetrahedral case, the sites are considered to be equivalent and connected by the same effective coupling. Following Refs.~\cite{VONOERTZEN200643,PRC-16O_QPT}, the reduced effective Hamiltonian is written as

\begin{equation}
H_{jj}=\epsilon, \qquad H_{jk}=-\mu, \qquad (j\neq k).
\end{equation}

Here, $\epsilon$ is the common on-site energy and $-\mu$ represents the effective coupling between the four cluster centres, with $\mu>0$. In this work, we obtain $\mu$ from the selected RHB levels. The resulting H\"uckel spectrum contains one lower fully symmetric bonding-like state, $E_s$, and three raised orthogonal partner states, $E_{\perp}$, which are degenerate in the ideal tetrahedral limit:

\begin{equation}
E_s=\epsilon-3\mu,
\qquad
E_{\perp}=\epsilon+\mu .
\end{equation}

The energy splitting is therefore

\begin{equation}
\Delta E = E_{\perp}-E_s=4\mu,
\end{equation}

so that

\begin{equation}
\mu=\frac{E_{\perp}-E_s}{4},
\qquad
\epsilon=\frac{E_s+3E_{\perp}}{4}=E_{\perp}-\mu .
\end{equation}
\begin{figure}[htbp!]
\centering
\includegraphics[width=\linewidth]{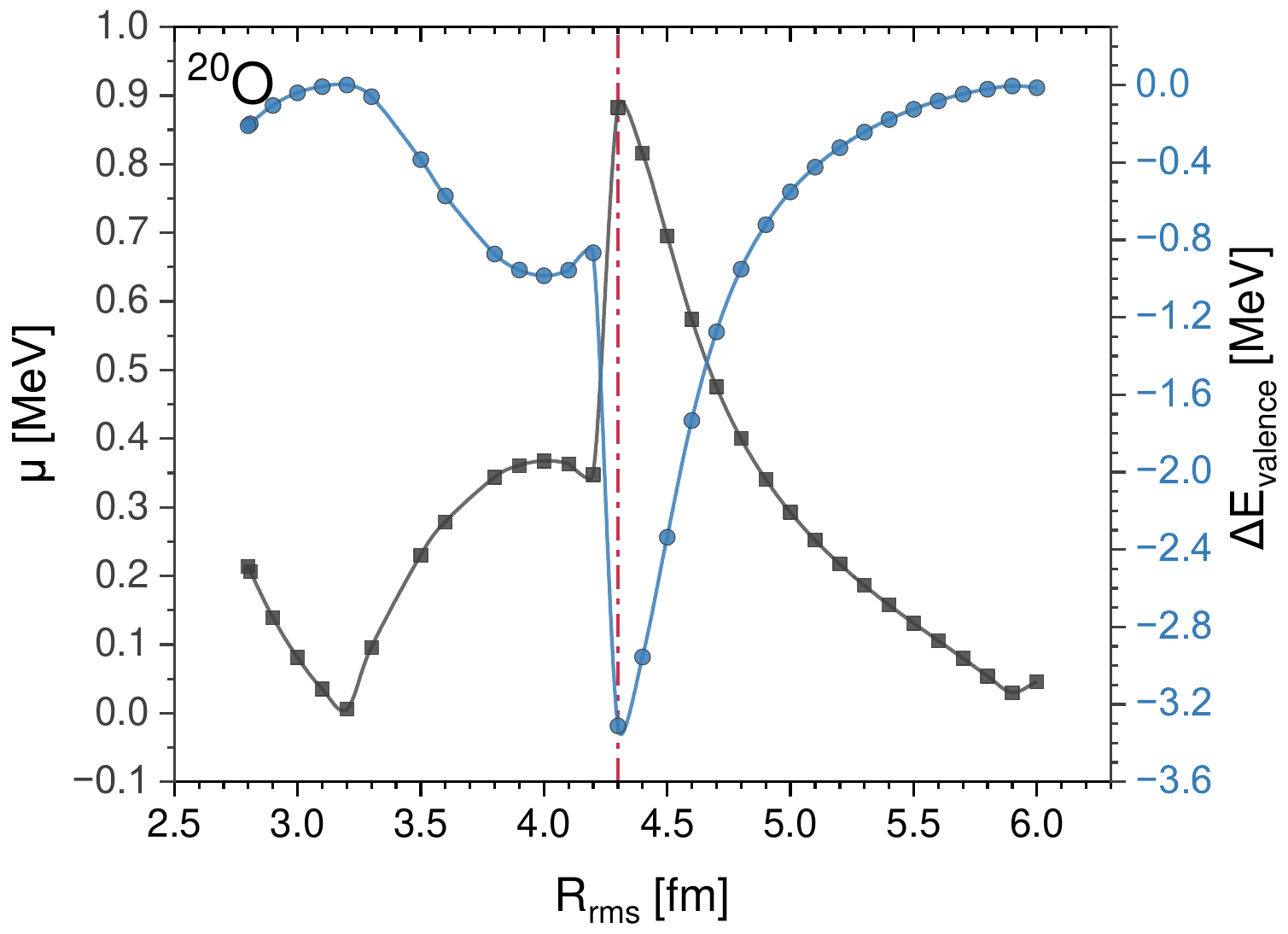}
\caption{Effective coupling $\mu$ and occupation-weighted H\"uckel energy shift $\Delta E_{\mathrm{val}}$ relative to the four-level centroid as functions of the constrained $R_{\mathrm{rms}}$ for $^{20}$O. Gray squares denote $\mu$ on the left-hand axis, and blue circles denote $\Delta E_{\mathrm{val}}$ on the right-hand axis. The transition radius $R_{\mathrm{PT}}$ is indicated by the red dashed-dotted line.}
\label{fig:mu-DeltaEVal}
\end{figure}

The self-consistent RHB calculations include spin–orbit coupling and are not constrained to an exact tetrahedral four-site geometry. Nonetheless, the three selected upper canonical levels remain approximately degenerate in each case and consistent with the threefold upper-level structure of the ideal model. Therefore, $E_\perp$ is taken as the centroid of the selected upper partner orbitals. When the centroid energy, $\epsilon$, is subtracted, the H\"uckel eigenvalues become $(-3\mu, +\mu, +\mu, +\mu)$, which add up to zero. In the ideal degeneracy limit, the H\"uckel single-particle energy shift relative to the four-level centroid $\epsilon$ is given by

\begin{equation}
\Delta E_{\mathrm{val}}=n_s(-3\mu)+n_{\perp}(+\mu).
\end{equation}

Here, $n_s$ denotes the occupation number of the lower fully symmetric molecular orbital, while $n_\perp$ is the summed occupation of the three upper partner orbitals. The occupation factors are necessary here since the H\"uckel energy shift is a many-neutron quantity. Each neutron occupying the lower state contributes $-3\mu$, whereas each neutron occupying an upper partner state contributes $+\mu$. In the RHB calculation, the occupation of a canonical orbital is obtained from its pairing occupation probability as $n_i=2v_i^2$. Here, the factor of two accounts for the time-reversed Kramers pair. 

\begin{equation}
n_s=2v_s^2, \quad \mathrm{and} \quad n_\perp=2\sum_{\alpha\in\perp}v_\alpha^2 .
\end{equation}

For an unpaired, fully occupied lower orbital, $n_s=2$. Fractional values of $v_i^2$ arise primarily from pairing correlations in the RHB solution. With this mapping, we calculate $\mu$ and $\Delta E_{\mathrm{val}}$ and in figure~\ref{fig:mu-DeltaEVal}, we show how they change with the constrained $R_{\mathrm{rms}}$ for $^{20}$O. It is seen that the parameter $\mu$ reaches its largest value in the transition region, where the tetrahedral cluster configuration appears. Before this point, $\mu$ should be regarded only as an effective spectral-splitting parameter since a well-defined cluster is not yet established. As we continue to expand the nucleus, the $\mu$ value decreases sharply and approaches zero. This shows that effective effective coupling between the four cluster centres decreases as the $\alpha$-like substructures move farther apart. As $R_{\mathrm{rms}}$ increases, the magnitude of $\Delta E_{\mathrm{val}}$ also gets closer to zero. This shows that occupation-weighted H\"uckel energy shift of the selected tetrahedral multiplet becomes smaller. We should note that the decrease in $\mu$ and $\Delta E_{\mathrm{val}}$ does not mean that the tetrahedral $4\alpha$-like structure disappears. The clustered configuration is already shown by density localisation and a non-zero $\beta_{32}$ deformation. In contrast, $\mu$ and $\Delta E_{\mathrm{val}}$ only measure the molecular contribution of the chosen valence neutrons, based on the centroid. When $R_{\mathrm{rms}}$ is large, their lower values simply show that effective coupling between the four cluster centres becomes weaker and eventually approaches to zero as the $\alpha$-like substructures move apart.

We obtain similar results for $^{18}$O and $^{24}$O (see Supplemental Material). The results show systematic isotope dependence at the respective localisation points. Just after the transition point, the value of $\mu$ decreases from approximately $1.0$~MeV in $^{18}$O to $0.9$~MeV in $^{20}$O and $0.7$~MeV in $^{24}$O. At the same time, $\Delta E_{\mathrm{val}}$ changes from $-5.9$~MeV in $^{18}$O to $-3.3$~MeV in $^{20}$O, and $-0.8$~MeV in $^{24}$O. In the H\"uckel mapping, $\Delta E_{\mathrm{val}}$ becomes less negative as more neutrons fill the upper partner states and $\mu$ goes down. When an upper state is filled, it adds $+\mu$ compared to the centroid. This addition helps balance the $-3\mu$ from each neutron in the lower symmetric state. The $\Delta E_{\mathrm{val}}$ values apply only to the chosen tetrahedral multiplet and do not show the total binding energy for valence neutrons.

\section{Summary and conclusions}\label{sect:conc}
In this study, we investigate the impact of neutron excess on density-driven localisation in neutron-rich oxygen isotopes under radial expansion of nuclei. The calculations are performed within the multidimensionally constrained relativistic Hartree--Bogoliubov framework using the DD-ME2 covariant energy-density functional. To obtain dilute configurations, we constrain the mass rms radius, $R_{\mathrm{rms}}$, to larger values up to 6.0 fm, suppress quadrupole deformation, and allow all other symmetry-allowed deformation degrees of freedom to evolve self-consistently.

For all oxygen isotopes, our calculations show a Mott-like structural rearrangement from a spatially delocalised nuclear configuration to a tetrahedral $4\alpha$-like configuration. The transition region is associated with the development of a non-zero $\beta_{32}$ deformation, sharp changes in the neutron pairing energy, and the reorganisation of the proton and neutron densities. We find that excess neutrons shift the onset of tetrahedral deformation and development of localisation to larger values of the total $R_{\mathrm{rms}}$, but do not suppress it. Furthermore, transition occurs at similar values of the relative dilution parameter, $\chi_{\mathrm{PT}}\approx 0.28-0.31$, for all nuclei. This suggests that the critical relative dilution remains nearly isotope independent and that the overall dilution of the nucleus drives the localisation of the underlying $^{16}$O-like subsystem into a $4\alpha$-like core, while the excess neutrons in $^{18}$O, $^{20}$O, and $^{24}$O primarily modify the density distribution of the resulting clustered configuration.  
The three-dimensional density distributions show that the valence neutrons remain more spatially delocalised than the $4\alpha$-like core and modify the overall density profile by increasing the density in the central region and between the clusters. Hence, they increase the spatial connectivity between the $\alpha$-like substructures and smooths the density profile.

A phenomenological four-centre H\"uckel-type mapping is used to interpret the selected valence neutron multiplet in terms of molecular-orbital-like states. The net occupation-weighted H\"uckel energy shift, $\Delta E_{\mathrm{val}}$, relative to the four-level centroid depends on the occupations of the lower fully symmetric bonding-like state and the upper partner states. Just after the localisation, the effective coupling $\mu$ decreases, while increasing occupation of the upper partner states with neutron number partly offsets the negative contribution of the lower state, making $\Delta E_{\mathrm{val}}$ less negative. Together with the behaviour of valence neutron density, these trends are consistent with a molecular-orbital-like interpretation of the selected valence neutron states.

These findings motivate experimental studies of neutron-excess effects on clustering in oxygen isotopes. Direct comparison with spectra and decay observables requires calculations beyond the constrained configurations considered here.

\section*{Acknowledgements}
E.Y. acknowledges support from the UK STFC under award no. ST/Y000358/1.
This work used the DiRAC Data Intensive service
(DIaL3) at the University of Leicester, managed by
the University of Leicester Research Computing Ser
vice on behalf of the STFC DiRAC HPC Facility
(www.dirac.ac.uk). The DiRAC service at Leicester was funded by BEIS, UKRI and STFC capital funding and STFC operations grants. DiRAC is part of the UKRI Digital Research Infrastructure. This work used the DiRAC Complexity system, operated by the University of Leicester IT Services, which forms part of the STFC DiRAC HPC Facility (www.dirac.ac.uk ). This
equipment is funded by BIS National E-Infrastructure capital grant ST/K000373/1 and STFC DiRAC Operations grant ST/K0003259/1. DiRAC is part of the National e-Infrastructure.
\bibliographystyle{elsarticle-num}
\bibliography{apssamp}

@article{16O-hoyle-analogue,
  title = {$\ensuremath{\alpha}$-Particle Condensation in $^{16}\mathrm{O}$ Studied with a Full Four-Body Orthogonality Condition Model Calculation},
  author = {Funaki, Y. and Yamada, T. and Horiuchi, H. and R\"opke, G. and Schuck, P. and Tohsaki, A.},
  journal = {Phys. Rev. Lett.},
  volume = {101},
  issue = {8},
  pages = {082502},
  numpages = {4},
  year = {2008},
  month = {Aug},
  publisher = {American Physical Society},
  doi = {10.1103/PhysRevLett.101.082502},
  url = {https://link.aps.org/doi/10.1103/PhysRevLett.101.082502}
}

@article{PRC-16O_QPT,
  title = {$\ensuremath{\alpha}$-particle condensation: A nuclear quantum phase transition},
  author = {Ebran, J.-P. and Girod, M. and Khan, E. and Lasseri, R. D. and Schuck, P.},
  journal = {Phys. Rev. C},
  volume = {102},
  issue = {1},
  pages = {014305},
  numpages = {10},
  year = {2020},
  month = {Jul},
  publisher = {American Physical Society},
  doi = {10.1103/PhysRevC.102.014305},
  url = {https://link.aps.org/doi/10.1103/PhysRevC.102.014305}
}

@article{GIROD-clustering,
  title = {$\ensuremath{\alpha}$-Particle Clustering from Expanding Self-Conjugate Nuclei within the Hartree-Fock-Bogoliubov Approach},
  author = {Girod, M. and Schuck, P.},
  journal = {Phys. Rev. Lett.},
  volume = {111},
  issue = {13},
  pages = {132503},
  numpages = {5},
  year = {2013},
  month = {Sep},
  publisher = {American Physical Society},
  doi = {10.1103/PhysRevLett.111.132503},
  url = {https://link.aps.org/doi/10.1103/PhysRevLett.111.132503}
}

@article{dirhb,
title = {DIRHB—A relativistic self-consistent mean-field framework for atomic nuclei},
journal = {Computer Physics Communications},
volume = {185},
number = {6},
pages = {1808-1821},
year = {2014},
issn = {0010-4655},
doi = {https://doi.org/10.1016/j.cpc.2014.02.027},
url = {https://www.sciencedirect.com/science/article/pii/S0010465514000836},
author = {T. Nikšić and N. Paar and D. Vretenar and P. Ring},
}

@article{PhysRevC.81.015803,
  title = {Composition and thermodynamics of nuclear matter with light clusters},
  author = {Typel, S. and R\"opke, G. and Kl\"ahn, T. and Blaschke, D. and Wolter, H. H.},
  journal = {Phys. Rev. C},
  volume = {81},
  issue = {1},
  pages = {015803},
  numpages = {22},
  year = {2010},
  month = {Jan},
  publisher = {American Physical Society},
  doi = {10.1103/PhysRevC.81.015803},
  url = {https://link.aps.org/doi/10.1103/PhysRevC.81.015803}
}

@article{PhysRevC.71.024312,
  title = {New relativistic mean-field interaction with density-dependent meson-nucleon couplings},
  author = {Lalazissis, G. A. and Nik\ifmmode \check{s}\else \v{s}\fi{}i\ifmmode \acute{c}\else \'{c}\fi{}, T. and Vretenar, D. and Ring, P.},
  journal = {Phys. Rev. C},
  volume = {71},
  issue = {2},
  pages = {024312},
  numpages = {10},
  year = {2005},
  month = {Feb},
  publisher = {American Physical Society},
  doi = {10.1103/PhysRevC.71.024312},
  url = {https://link.aps.org/doi/10.1103/PhysRevC.71.024312}
}

@article{TIAN200944,
title = {A finite range pairing force for density functional theory in superfluid nuclei},
journal = {Physics Letters B},
volume = {676},
number = {1},
pages = {44-50},
year = {2009},
issn = {0370-2693},
doi = {https://doi.org/10.1016/j.physletb.2009.04.067},
url = {https://www.sciencedirect.com/science/article/pii/S0370269309004912},
author = {Y. Tian and Z.Y. Ma and P. Ring}
}

@article{PhysRevC.87.044307,
  title = {Localization and clustering in the nuclear Fermi liquid},
  author = {Ebran, J.-P. and Khan, E. and Nik\ifmmode \check{s}\else \v{s}\fi{}i\ifmmode \acute{c}\else \'{c}\fi{}, T. and Vretenar, D.},
  journal = {Phys. Rev. C},
  volume = {87},
  issue = {4},
  pages = {044307},
  numpages = {5},
  year = {2013},
  month = {Apr},
  publisher = {American Physical Society},
  doi = {10.1103/PhysRevC.87.044307},
  url = {https://link.aps.org/doi/10.1103/PhysRevC.87.044307}
}

@article{PhysRevC.90.054329,
  title = {Density functional theory studies of cluster states in nuclei},
  author = {Ebran, J.-P. and Khan, E. and Nik\ifmmode \check{s}\else \v{s}\fi{}i\ifmmode \acute{c}\else \'{c}\fi{}, T. and Vretenar, D.},
  journal = {Phys. Rev. C},
  volume = {90},
  issue = {5},
  pages = {054329},
  numpages = {11},
  year = {2014},
  month = {Nov},
  publisher = {American Physical Society},
  doi = {10.1103/PhysRevC.90.054329},
  url = {https://link.aps.org/doi/10.1103/PhysRevC.90.054329}
}

@article{FREER20141,
title = {The Hoyle state in 12C},
journal = {Progress in Particle and Nuclear Physics},
volume = {78},
pages = {1-23},
year = {2014},
issn = {0146-6410},
doi = {https://doi.org/10.1016/j.ppnp.2014.06.001},
url = {https://www.sciencedirect.com/science/article/pii/S0146641014000453},
author = {M. Freer and H.O.U. Fynbo}
}

@article{Freer_2007,
doi = {10.1088/0034-4885/70/12/R03},
url = {https://dx.doi.org/10.1088/0034-4885/70/12/R03},
year = {2007},
month = {nov},
publisher = {},
volume = {70},
number = {12},
pages = {2149},
author = {Martin Freer},
title = {The clustered nucleus—cluster structures in stable and unstable nuclei},
journal = {Reports on Progress in Physics}
}

@Article{Ebran2012,
author={Ebran, J.-P.
and Khan, E.
and Nik{\v{s}}i{\'{c}}, T.
and Vretenar, D.},
title={How atomic nuclei cluster},
journal={Nature},
year={2012},
month={Jul},
day={01},
volume={487},
number={7407},
pages={341-344},
issn={1476-4687},
doi={10.1038/nature11246},
url={https://doi.org/10.1038/nature11246}
}

@article{ROPKE1982536,
title = {Particle clustering and Mott transitions in nuclear matter at finite temperature: (I). Method and general aspects},
journal = {Nuclear Physics A},
volume = {379},
number = {3},
pages = {536-552},
year = {1982},
issn = {0375-9474},
doi = {https://doi.org/10.1016/0375-9474(82)90013-6},
url = {https://www.sciencedirect.com/science/article/pii/0375947482900136},
author = {G. Röpke and L. Münchow and H. Schulz}
}

@article{RevModPhys.89.011002,
  title = {Colloquium: Status of $\ensuremath{\alpha}$-particle condensate structure of the Hoyle state},
  author = {Tohsaki, Akihiro and Horiuchi, Hisashi and Schuck, Peter and R\"opke, Gerd},
  journal = {Rev. Mod. Phys.},
  volume = {89},
  issue = {1},
  pages = {011002},
  numpages = {16},
  year = {2017},
  month = {Jan},
  publisher = {American Physical Society},
  doi = {10.1103/RevModPhys.89.011002},
  url = {https://link.aps.org/doi/10.1103/RevModPhys.89.011002}
}

@Article{Staszczak2010,
author={Staszczak, A.
and Stoitsov, M.
and Baran, A.
and Nazarewicz, W.},
title={Augmented Lagrangian method for constrained nuclear density functional theory},
journal={The European Physical Journal A},
year={2010},
month={Oct},
day={01},
volume={46},
number={1},
pages={85-90},
issn={1434-601X},
doi={10.1140/epja/i2010-11018-9},
url={https://doi.org/10.1140/epja/i2010-11018-9}
}

@book{ring1980nuclear,
  title={The Nuclear Many-body Problem},
  author={Ring, P. and Schuck, P.},
  isbn={9783540098201},
  lccn={79025447},
  year={1980},
  publisher={Springer},
  address={Heidelberg}
}

@article{VRETENAR2005101,
title = {Relativistic Hartree–Bogoliubov theory: static and dynamic aspects of exotic nuclear structure},
journal = {Physics Reports},
volume = {409},
number = {3},
pages = {101-259},
year = {2005},
issn = {0370-1573},
doi = {https://doi.org/10.1016/j.physrep.2004.10.001},
url = {https://www.sciencedirect.com/science/article/pii/S0370157304004545},
author = {D. Vretenar and A.V. Afanasjev and G.A. Lalazissis and P. Ring}
}

@article{PhysRevLett.80.3177,
  title = {Four-Particle Condensate in Strongly Coupled Fermion Systems},
  author = {R\"opke, G. and Schnell, A. and Schuck, P. and Nozi\`eres, P.},
  journal = {Phys. Rev. Lett.},
  volume = {80},
  issue = {15},
  pages = {3177--3180},
  numpages = {0},
  year = {1998},
  month = {Apr},
  publisher = {American Physical Society},
  doi = {10.1103/PhysRevLett.80.3177},
  url = {https://link.aps.org/doi/10.1103/PhysRevLett.80.3177}
}

@article{CHEN20231119,
title = {New evidence of the Hoyle-like structure in 16O},
journal = {Science Bulletin},
volume = {68},
number = {11},
pages = {1119-1126},
year = {2023},
issn = {2095-9273},
doi = {https://doi.org/10.1016/j.scib.2023.04.031},
url = {https://www.sciencedirect.com/science/article/pii/S2095927323002852},
author = "{J. Chen, Y. Ye, K. Ma, \textit{et al}.}"
}

@article{DAVIES-16O-PT-FT,
  title = {$\ensuremath{\alpha}$-particle condensation in diluted $^{16}\mathrm{O}$ at finite temperature},
  author = {Davies, M. and Y\"uksel, E. and Ebran, J.-P. and Khan, E. and Stevenson, P.},
  journal = {Phys. Rev. C},
  volume = {112},
  issue = {1},
  pages = {014311},
  numpages = {9},
  year = {2025},
  month = {Jul},
  publisher = {American Physical Society},
  doi = {10.1103/wrzp-c3n9},
  url = {https://link.aps.org/doi/10.1103/wrzp-c3n9}
}

@article{VONOERTZEN2003465,
title = {Covalently bound molecular states in beryllium and carbon isotopes},
journal = {Comptes Rendus Physique},
volume = {4},
number = {4},
pages = {465-474},
year = {2003},
issn = {1631-0705},
doi = {https://doi.org/10.1016/S1631-0705(03)00052-5},
url = {https://www.sciencedirect.com/science/article/pii/S1631070503000525},
author = {Wolfram {von Oertzen} and Hans-Gerhard Bohlen},
}

@article{VONOERTZEN1997,
journal = {Z Phys A},
number = {357},
pages = {355-365},
year = {1997},
issn = {1631-0705},
doi = {https://doi.org/10.1007/s002180050255},
author = {Wolfram {von Oertzen}},
}

@book{Clusters_in_nuclei,
    author = {C. Beck},
    title = {Clusters in Nuclei},
    subtitle = {Vol. 1},
    publisher = {Springer Berlin, Heidelberg},
    year = {2010}
}

@article{PhysRevC.93.024611,
  title = {New experimental investigation of the structure of $^{10}\mathrm{Be}$ and $^{16}\mathrm{C}$ by means of intermediate-energy sequential breakup},
  author = {Dell'Aquila, D. and Lombardo, I. and Acosta, L. and Andolina, R. and Auditore, L. and Cardella, G. and Chatterjiee, M. B. and De Filippo, E. and Francalanza, L. and Gnoffo, B. and Lanzalone, G. and Pagano, A. and Pagano, E. V. and Papa, M. and Pirrone, S. and Politi, G. and Porto, F. and Quattrocchi, L. and Rizzo, F. and Rosato, E. and Russotto, P. and Trifir\`o, A. and Trimarchi, M. and Verde, G. and Vigilante, M.},
  journal = {Phys. Rev. C},
  volume = {93},
  issue = {2},
  pages = {024611},
  numpages = {8},
  year = {2016},
  month = {Feb},
  publisher = {American Physical Society},
  doi = {10.1103/PhysRevC.93.024611},
  url = {https://link.aps.org/doi/10.1103/PhysRevC.93.024611}
}

@article{Di_Pietro_2018,
doi = {10.1088/1742-6596/966/1/012040},
url = {https://doi.org/10.1088/1742-6596/966/1/012040},
year = {2018},
month = {feb},
publisher = {IOP Publishing},
volume = {966},
number = {1},
pages = {012040},
author = {Di Pietro, A. and Fernández-García, J. P. and Ferrera, F. and Figuera, P. and Fisichella, M. and Lattuada, M. and Marletta, S. and Marchetta, C. and Torresi, D. and Alcorta, M. and Borge, M. J. G. and Davinson, T. and Heinitz, S. and Laird, A. M. and Shotter, A. C. and Schumann, D. and Soic, N. and Tengblad, O. and Zadro, M.},
title = {Experimental investigation of exotic clustering in 13B and 14C using the resonance scattering method},
journal = {Journal of Physics: Conference Series}
}

@article{PhysRevC.70.064607,
  title = {Measurements of the breakup and neutron removal cross sections for $^{16}\mathrm{C}$},
  author = {Ashwood, N. I. and Freer, M. and Ang\'elique, J. C. and Bouchat, V. and Catford, W. N. and Clarke, N. M. and Curtis, N. and Dorvaux, O. and Hanappe, F. and Kerckx, Y. and Labiche, M. and Lecouey, J. L. and Marqu\'es, F. M. and Materna, T. and Ninane, A. and Normand, G. and Orr, N. A. and Pain, S. and Soi\ifmmode \acute{c}\else \'{c}\fi{}, N. and Stuttg\'e, L. and Timis, C. and Unshakova, A. and Ziman, V. A.},
  journal = {Phys. Rev. C},
  volume = {70},
  issue = {6},
  pages = {064607},
  numpages = {5},
  year = {2004},
  month = {Dec},
  publisher = {American Physical Society},
  doi = {10.1103/PhysRevC.70.064607},
  url = {https://link.aps.org/doi/10.1103/PhysRevC.70.064607}
}

@article{PhysRevLett.124.192501,
  title = {Positive-Parity Linear-Chain Molecular Band in $^{16}\mathrm{C}$},
  author = {Liu, Y. and Ye, Y. L. and Lou, J. L. and Yang, X. F. and Baba, T. and Kimura, M. and Yang, B. and Li, Z. H. and Li, Q. T. and Xu, J. Y. and Ge, Y. C. and Hua, H. and Wang, J. S. and Yang, Y. Y. and Ma, P. and Bai, Z. and Hu, Q. and Liu, W. and Ma, K. and Tao, L. C. and Jiang, Y. and Hu, L. Y. and Zang, H. L. and Feng, J. and Wu, H. Y. and Han, J. X. and Bai, S. W. and Li, G. and Yu, H. Z. and Huang, S. W. and Chen, Z. Q. and Sun, X. H. and Li, J. J. and Tan, Z. W. and Gao, Z. H. and Duan, F. F. and Tan, J. H. and Sun, S. Q. and Song, Y. S.},
  journal = {Phys. Rev. Lett.},
  volume = {124},
  issue = {19},
  pages = {192501},
  numpages = {6},
  year = {2020},
  month = {May},
  publisher = {American Physical Society},
  doi = {10.1103/PhysRevLett.124.192501},
  url = {https://link.aps.org/doi/10.1103/PhysRevLett.124.192501}
}

@article{VONOERTZEN1996,
    author = {W. von Oertzen},
    issn = {0939-7922},
    journal = {Z. Phys.},
    number = {1},
    title = {Two-center molecular states in 9B, 9Be, 10Be, and 10B},
    volume = {354},
    year = {1996},
}

@article{hoyle1954nuclear,
  title={On Nuclear Reactions Occuring in Very Hot STARS. I. the Synthesis of Elements from Carbon to Nickel.},
  author={Hoyle, Fred},
  journal={Astrophysical Journal Supplement, vol. 1, p. 121 (1954)},
  volume={1},
  pages={121},
  year={1954}
}

@article{PhysRevC.90.064319,
  title = {$3\ensuremath{\alpha}$ clustering in excited states of $^{16}\mathrm{C}$},
  author = {Baba, T. and Chiba, Y. and Kimura, M.},
  journal = {Phys. Rev. C},
  volume = {90},
  issue = {6},
  pages = {064319},
  numpages = {7},
  year = {2014},
  month = {Dec},
  publisher = {American Physical Society},
  doi = {10.1103/PhysRevC.90.064319},
  url = {https://link.aps.org/doi/10.1103/PhysRevC.90.064319}
}

@article{PhysRevC.97.054315,
  title = {Characteristic $\ensuremath{\alpha}$ and $^{6}\mathrm{He}$ decays of linear-chain structures in $^{16}\mathrm{C}$},
  author = {Baba, T. and Kimura, M.},
  journal = {Phys. Rev. C},
  volume = {97},
  issue = {5},
  pages = {054315},
  numpages = {12},
  year = {2018},
  month = {May},
  publisher = {American Physical Society},
  doi = {10.1103/PhysRevC.97.054315},
  url = {https://link.aps.org/doi/10.1103/PhysRevC.97.054315}
}

@article{PhysRevC.64.014301,
  title = {Molecular-orbital structure in neutron-rich C isotopes},
  author = {Itagaki, N. and Okabe, S. and Ikeda, K. and Tanihata, I.},
  journal = {Phys. Rev. C},
  volume = {64},
  issue = {1},
  pages = {014301},
  numpages = {5},
  year = {2001},
  month = {May},
  publisher = {American Physical Society},
  doi = {10.1103/PhysRevC.64.014301},
  url = {https://link.aps.org/doi/10.1103/PhysRevC.64.014301}
}

@article{Huckel1931,
  title = {`Quantentheoretische Beitr\"age zum Benzolproblem},
  author = {E. H\"uckel},
  journal = {Zeitschrift f\"ur Physik},
  volume = {70},
  issue = {3},
  pages = {204-286},
  year = {1931},
  doi = {10.1007/BF01339530},
  url = {https://doi.org/10.1007/BF01339530}
}

@article{Huckel1932,
  title = {Quantentheoretische Beitr\"age zum Problem der aromatischen und unges\"attigten Verbindungen. III},
  author = {E. H\"uckel},
  journal = {Zeitschrift f\"ur Physik},
  volume = {76},
  issue = {9},
  pages = {628-648},
  year = {1932},
  doi = {10.1007/BF01341936},
  url = {https://doi.org/10.1007/BF01341936}
}

@article{Ebran2017,
doi = {10.1088/1361-6471/aa809b},
url = {https://doi.org/10.1088/1361-6471/aa809b},
year = {2017},
month = {aug},
publisher = {IOP Publishing},
volume = {44},
number = {10},
pages = {103001},
author = {Ebran, J-P and Khan, E and Nikšić, T and Vretenar, D},
title = {Localization and clustering in atomic nuclei},
journal = {Journal of Physics G: Nuclear and Particle Physics},
}

@article{VONOERTZEN2004,
  author  = {W. von Oertzen},
  title   = {Nuclear Structure with Clusters and Covalent Valence Neutrons},
  journal = {Nuclear Physics A},
  volume  = {734},
  pages   = {385--394},
  year    = {2004},
  doi     = {10.1016/j.nuclphysa.2004.01.073}
}

@article{Brooks_2025,
doi = {10.1088/1361-6471/adf5d9},
url = {https://doi.org/10.1088/1361-6471/adf5d9},
year = {2025},
month = {aug},
publisher = {IOP Publishing},
volume = {52},
number = {8},
pages = {085104},
author = {Brooks, A D and Bishop, J and Kokalova, Tz and Pirrie, S and Wheldon, C},
title = {Application of a molecular model to band-head states in 9B},
journal = {Journal of Physics G: Nuclear and Particle Physics},
}

@article{VONOERTZEN200643,
title = {Nuclear clusters and nuclear molecules},
journal = {Physics Reports},
volume = {432},
number = {2},
pages = {43-113},
year = {2006},
issn = {0370-1573},
doi = {https://doi.org/10.1016/j.physrep.2006.07.001},
url = {https://www.sciencedirect.com/science/article/pii/S0370157306002626},
author = {W. {von Oertzen} and Martin Freer and Yoshiko Kanada-En’yo}
}

@article{BEYER2000247,
title = {The $\alpha$-particle in nuclear matter},
journal = {Physics Letters B},
volume = {488},
number = {3},
pages = {247-253},
year = {2000},
issn = {0370-2693},
doi = {https://doi.org/10.1016/S0370-2693(00)00908-4},
url = {https://www.sciencedirect.com/science/article/pii/S0370269300009084},
author = {M Beyer and S.A Sofianos and C Kuhrts and G Röpke and P Schuck}
}

@article{Ropke2015,
  author  = {G. R{\"o}pke},
  title   = {Clusters in Nuclear Matter and Mott Points},
  journal = {Physics of Particles and Nuclei},
  volume  = {46},
  number  = {5},
  pages   = {772--776},
  year    = {2015},
  doi     = {10.1134/S106377961505024X},
  url     = {https://doi.org/10.1134/S106377961505024X}
}

@article{PhysRevLett.112.152501,
  title = {Evidence for Tetrahedral Symmetry in $^{16}\mathrm{O}$},
  author = {Bijker, R. and Iachello, F.},
  journal = {Phys. Rev. Lett.},
  volume = {112},
  issue = {15},
  pages = {152501},
  numpages = {5},
  year = {2014},
  month = {Apr},
  publisher = {American Physical Society},
  doi = {10.1103/PhysRevLett.112.152501},
  url = {https://link.aps.org/doi/10.1103/PhysRevLett.112.152501}
}

@article{PhysRevLett.112.102501,
  title = {Ab Initio Calculation of the Spectrum and Structure of $^{16}\mathrm{O}$},
  author = {Epelbaum, Evgeny and Krebs, Hermann and L\"ahde, Timo A. and Lee, Dean and Mei\ss{}ner, Ulf-G. and Rupak, Gautam},
  journal = {Phys. Rev. Lett.},
  volume = {112},
  issue = {10},
  pages = {102501},
  numpages = {5},
  year = {2014},
  month = {Mar},
  publisher = {American Physical Society},
  doi = {10.1103/PhysRevLett.112.102501},
  url = {https://link.aps.org/doi/10.1103/PhysRevLett.112.102501}
}

@article{10.1143/PTP.119.403,
    author = {Furutachi, Naoya and Kimura, Masaaki and Doté, Akinobu and Kanada-En'yo, Yoshiko and Oryu, Shinsho},
    title = {Cluster Structures in Oxygen Isotopes},
    journal = {Progress of Theoretical Physics},
    volume = {119},
    number = {3},
    pages = {403-420},
    year = {2008},
    month = {03},
    issn = {0033-068X},
    doi = {10.1143/PTP.119.403},
    url = {https://doi.org/10.1143/PTP.119.403},
    eprint = {https://academic.oup.com/ptp/article-pdf/119/3/403/5296008/119-3-403.pdf},
}

@article{PhysRevC.25.1108,
  title = {Test of tetrahedral symmetry in the $^{16}\mathrm{O}$ nucleus},
  author = {Robson, D.},
  journal = {Phys. Rev. C},
  volume = {25},
  issue = {2},
  pages = {1108(R)--1110(R)},
  numpages = {0},
  year = {1982},
  month = {Feb},
  publisher = {American Physical Society},
  doi = {10.1103/PhysRevC.25.1108},
  url = {https://link.aps.org/doi/10.1103/PhysRevC.25.1108}
}

@article{BRINK1970143,
title = {Investigation of the alpha-particle model for light nuclei},
journal = {Physics Letters B},
volume = {33},
number = {2},
pages = {143-146},
year = {1970},
issn = {0370-2693},
doi = {https://doi.org/10.1016/0370-2693(70)90284-4},
url = {https://www.sciencedirect.com/science/article/pii/0370269370902844},
author = {D.M. Brink and H. Friedrich and A. Weiguny and C.W. Wong}
}

@article{PhysRevLett.42.876,
  title = {Evidence for the Tetrahedral Nature of $^{16}\mathrm{O}$},
  author = {Robson, D.},
  journal = {Phys. Rev. Lett.},
  volume = {42},
  issue = {14},
  pages = {876--879},
  numpages = {0},
  year = {1979},
  month = {Apr},
  publisher = {American Physical Society},
  doi = {10.1103/PhysRevLett.42.876},
  url = {https://link.aps.org/doi/10.1103/PhysRevLett.42.876}
}

@article{v1938neuere,
  title={Neuere Modellvorstellungen {\"u}ber den Bau der Atomkerne},
  author={v. Weizs{\"a}cker, CF},
  journal={Naturwissenschaften},
  volume={26},
  number={14},
  pages={209--217},
  year={1938},
  publisher={Springer}
}

@article{wheeler1937mathematical,
  title={On the mathematical description of light nuclei by the method of resonating group structure},
  author={Wheeler, John A},
  journal={Physical Review},
  volume={52},
  number={11},
  pages={1107},
  year={1937},
  publisher={APS}
}

@article{PhysRevC.99.064315,
  title = {Investigation of the $^{14}\mathrm{C}+\ensuremath{\alpha}$ molecular configuration in $^{18}\mathrm{O}$ by means of transfer and sequential decay reaction},
  author = {Yang, B. and Ye, Y. L. and Feng, J. and Lin, C. J. and Jia, H. M. and Li, Z. H. and Lou, J. L. and Li, Q. T. and Ge, Y. C. and Yang, X. F. and Hua, H. and Li, J. and Zang, H. L. and Liu, Q. and Jiang, W. and Li, C. G. and Liu, Y. and Chen, Z. Q. and Wu, H. Y. and Wang, C. G. and Liu, W. and Wang, X. and Li, J. J. and Luo, D. W. and Jiang, Y. and Bai, S. W. and Xu, J. Y. and Ma, N. R. and Sun, L. J. and Wang, D. X. and Yang, Z. H. and Chen, J.},
  journal = {Phys. Rev. C},
  volume = {99},
  issue = {6},
  pages = {064315},
  numpages = {9},
  year = {2019},
  month = {Jun},
  publisher = {American Physical Society},
  doi = {10.1103/PhysRevC.99.064315},
  url = {https://link.aps.org/doi/10.1103/PhysRevC.99.064315}
}

@article{PhysRevC.90.024327,
  title = {$\ensuremath{\alpha}$-cluster structure of $^{18}\mathrm{O}$},
  author = {Avila, M. L. and Rogachev, G. V. and Goldberg, V. Z. and Johnson, E. D. and Kemper, K. W. and Tchuvil'sky, Yu. M. and Volya, A. S.},
  journal = {Phys. Rev. C},
  volume = {90},
  issue = {2},
  pages = {024327},
  numpages = {19},
  year = {2014},
  month = {Aug},
  publisher = {American Physical Society},
  doi = {10.1103/PhysRevC.90.024327},
  url = {https://link.aps.org/doi/10.1103/PhysRevC.90.024327}
}

@article{PhysRevC.78.034318,
  title = {Relativistic nuclear energy density functionals: Adjusting parameters to binding energies},
  author = {Nik\ifmmode \check{s}\else \v{s}\fi{}i\ifmmode \acute{c}\else \'{c}\fi{}, T. and Vretenar, D. and Ring, P.},
  journal = {Phys. Rev. C},
  volume = {78},
  issue = {3},
  pages = {034318},
  numpages = {19},
  year = {2008},
  month = {Sep},
  publisher = {American Physical Society},
  doi = {10.1103/PhysRevC.78.034318},
  url = {https://link.aps.org/doi/10.1103/PhysRevC.78.034318}
}
\end{document}